\RequirePackage{lineno}
\documentclass[twocolumn,aps,prd,groupedaddress,nofootinbib,showpacs]{revtex4}
\usepackage{graphicx}
\usepackage{amsmath}
\usepackage{bm}
\usepackage{slashed}
\usepackage{epsfig}
\usepackage{amsfonts}

\newcommand\gev{\mathrm{~GeV}}

\newcommand{\jpsi}{{J/\psi}}

\newcommand{\state}[4]{{^{#1}\hspace{-0.6mm}#2_{#3}^{[#4]}}}

\newcommand\CScSa{\state{3}{S}{1}{1}}

\newcommand\COaSz{\state{1}{S}{0}{8}}

\newcommand\COcSa{\state{3}{S}{1}{8}}
\newcommand\COcPz{\state{3}{P}{0}{8}}

\newcommand\COcPj{\state{3}{P}{J}{8}}

\newcommand\mo{{\mathcal O}}

\newcommand{\LDME}[2]{\langle\mo^{#1}(#2)\rangle}
\newcommand\mops{\LDME{\jpsi}{\CScSa}}
\newcommand\mopa{\LDME{\jpsi}{\COaSz}}
\newcommand\mopb{\LDME{\jpsi}{\COcSa}}
\newcommand\mopc{\LDME{\jpsi}{\COcPz}}

\newcommand{\vt}[1]{{{\boldsymbol #1}_\perp}}
\newcommand{\vtn}[2]{{{\boldsymbol #1}_{#2\perp}}}

\newcommand{\vp}{{\vt{p}}}
\newcommand{\vk}{{\vt{k}}}
\newcommand{\vka}{{\vtn{k}{1}}}

\newcommand{\vtp}[1]{{{\boldsymbol #1}'_\perp}}

\newcommand{\vkp}{{\vtp{k}}}

\def\be{\begin{equation}}
\def\ee{\end{equation}}
\def\bea{\begin{eqnarray}}
\def\eea{\end{eqnarray}}

\def\gev{\mathrm{~GeV}}

\begin{document}


\title{$J/\psi$ production and suppression in high energy proton-nucleus collisions}


\author{Yan-Qing Ma$^{1,2}$}
\author{Raju Venugopalan$^{3}$}
\author{Hong-Fei Zhang$^{4}$}
\affiliation{
$^{1}$ Maryland Center for Fundamental Physics, University of Maryland, College Park, Maryland 20742, USA.\\
$^{2}$ Center for High Energy Physics, Peking
University, Beijing 100871, China.\\
$^{3}$ Physics Department, Brookhaven National Laboratory, Upton, New York 11973-5000, USA. \\
$^{4}$ Department of Physics, School of Biomedical Engineering, Third Military Medical University, Chongqing 400038, China.
}%
\date{\today}

\begin{abstract}
We apply a Color Glass Condensate+Non-Relativistic QCD (CGC+NRQCD) framework to compute $J/\psi$ production in deuteron-nucleus collisions at RHIC and proton-nucleus collisions at the LHC.
Our results match smoothly at high $p_\perp$ to a next-to-leading order perturbative QCD + NRQCD computation.
Excellent agreement is obtained for $p_\perp$ spectra at RHIC and LHC for central and forward rapidities,
as well as for the normalized ratio $R_{pA}$ of these results to spectra in proton-proton collisions.
In particular, we observe that the $R_{pA}$ data is strongly bounded by our computations of the same for each of the individual NRQCD channels;
this result provides strong evidence that our description is robust against uncertainties in initial conditions and hadronization mechanisms.
\end{abstract}


\maketitle
The copious production of heavy quarkonium states at high energy colliders has inaugurated a new era of precision studies of such states~\cite{Brambilla:2010cs}. In proton-proton collisions (p+p), next-to-leading order (NLO) perturbative studies are available \cite{Butenschoen:2012px,Chao:2012iv,Gong:2012ug,Shao:2014yta} within the Non-Relativistic QCD (NRQCD) factorization framework~\cite{Bodwin:1994jh}. These computations can be further improved by employing QCD factorization \cite{Kang:2014tta,Kang:2014pya} to resum large logarithms $\ln(p_\perp/M)$ in the ratio of the transverse momentum $p_\perp$ to the quark mass $M$.
A comparison of these studies with collider data therefore provides key insight into the formation and hadronization of heavy quark-antiquark pair ($Q\bar{Q}$-pair) states in QCD.

In proton-nucleus (p+A) collisions, additional features of $Q\bar{Q}$-pair production and hadronization can be tested. These include many-body QCD effects such as multiple scattering and shadowing of gluon distributions in nuclei, as well as the radiative energy loss induced in the scattering of the $Q\bar{Q}$-pair off the colored medium. Besides these insights into many-body QCD dynamics,  p+A collisions also  provide an important benchmark for understanding the interactions of heavy quarks in the hot and dense medium created in heavy ion collisions.

For small gluon momentum fractions $x$, their distributions saturate with a dynamically generated saturation scale $Q_S(x)$ \cite{Gribov:1984tu,Mueller:1985wy,McLerran:1993ni,McLerran:1993ka}. This regime is accessed when  $p_\perp \lesssim Q_S$. The Color Glass Condensate (CGC) effective theory \cite{Iancu:2003xm,Gelis:2010nm} provides a quantitative framework to study many-body QCD effects in high energy scattering processes when $Q_S(x) >> \Lambda_{\rm QCD}$, where $\Lambda_{\rm QCD}$ is the fundamental QCD scale. In this limit, multiple scattering contributions can be absorbed into light like Wilson line correlators, which govern the shadowing and $p_\perp$ broadening of $Q\bar{Q}$-pair distributions at small $x$. Energy evolution of these correlations at small $x$ is described by the Balitsky-JIMWLK hierarchy of renormalization group equations~\cite{Balitsky:1995ub,Iancu:2000hn,JalilianMarian:1997dw}. Energy loss contributions, included in some models in the literature~\cite{Arleo:2012rs}, are formally NLO in the CGC framework~ \cite{Liou:2014rha}.

Expressions for $Q\bar{Q}$-pair production in p+A collisions in the CGC framework were derived previously in~\cite{Blaizot:2004wv,Fujii:2005vj,Fujii:2006ab,Kharzeev:2005zr,Kharzeev:2008nw,Dominguez:2011cy,Akcakaya:2012si}
as well as in related dipole approaches~\cite{Kopeliovich:2002yv,Motyka:2015kta}. For $p_\perp >> Q_S$, the results can be matched to those derived in perturbative QCD frameworks~\cite{Gelis:2003vh}.
In \cite{Fujii:2013gxa}, the matrix elements in the CGC framework were combined with the Color Evaporation hadronization Model (CEM) to compute $J/\psi$ production in proton-proton and proton (deuteron)-nucleus collisions at the LHC (RHIC)\footnote{For simplicity, we will generically call both sorts of collisions p+A collisions in the rest of the paper.}. The quantity
\begin{align}\label{eq:RpA}
R_{pA}=\frac{d\sigma_{pA}}{A\times d\sigma_{pp}}\,,
\end{align}
the ratio of the cross-sections in p+A collisions to p+p collisions, normalized by the atomic number $A$, was found to be suppressed relative to the data~\cite{Abelev:2013yxa,Aaij:2013zxa}. Very recently, the authors of \cite{Ducloue:2015gfa} argued that better agreement of the CGC+CEM model with the $R_{pA}$ data is obtained if nuclear effects were treated differently.  Here we shall apply NRQCD to describe the hadronization of $Q\bar{Q}$-pair and compute $J/\psi$ production in a CGC+NRQCD framework~\cite{Kang:2013hta}. In addition to providing a more systematic power counting, NRQCD allows one to smoothly match the CGC computations to successful NLO NRQCD computations  for $p_\perp >> Q_S$. This strategy was previously applied to successfully describe p+p collisions at RHIC and LHC~\cite{Ma:2014mri}.

For completeness, we outline the CGC+NRQCD formalism~\cite{Kang:2013hta,Ma:2014mri}. In NRQCD factorization, the production cross section of a quarkonium $H$ in the forward region of a p+A collision is expressed as \cite{Bodwin:1994jh}
\be \label{eq:NRQCD}
d\sigma^H_{pA}=\sum_\kappa d\hat{\sigma}_{pA}^\kappa\langle{\cal O}^H_\kappa\rangle\,,
\ee
where $\kappa=\state{2S+1}{L}{J}{c}$ denotes the quantum numbers of the intermediate $Q\bar{Q}$-pair in the standard spectroscopic notation for angular momentum. The superscript $c$ denotes the color state of the pair, which can be either color singlet (CS) with $c=1$ or color octet
(CO) with $c=8$. For $J/\psi$ production that will be studied here, the most important intermediate states are $\CScSa$, $\COaSz$, $\COcSa$ and $\COcPj$. In Eq.~\eqref{eq:NRQCD}, $\langle{\cal O}^H_\kappa\rangle$ are non-perturbative universal long distance matrix elements (LDMEs), which can be extracted from data, and $d\hat{\sigma}^\kappa$ are short-distance coefficients (SDCs) for the production of a $Q\bar{Q}$-pair, computed in perturbative QCD.

To calculate the SDCs in Eq.~\eqref{eq:NRQCD}, we apply the CGC effective field theory \cite{Gelis:2010nm,Blaizot:2004wv}, which results in the expressions~\cite{Kang:2013hta,Ma:2014mri},
\begin{align}\label{eq:dsktCS}
\begin{split}
\frac{d \hat{\sigma}_{pA}^\kappa}{d^2\vp d
y}\overset{\text{CS}}=&\frac{\alpha_s (\pi \bar{R}_A^2)}{(2\pi)^{9}
(N_c^2-1)} \underset{\vka,\vk,\vkp}{\int}
\frac{\varphi_{p,y_p}(\vka)}{k_{1\perp}^2}\\
&\hspace{-1.5cm}\times \mathcal{N}_{Y}(\vk)\mathcal{N}_{Y}(\vkp)\mathcal{N}_{Y}(\vp-\vka-\vk-\vkp)\,
{\cal G}^\kappa_1,
\end{split}
\end{align}
for the color-singlet $\CScSa$ channel, and
\begin{align}\label{eq:dsktCO}
\begin{split}
\frac{d \hat{\sigma}_{pA}^\kappa}{d^2\vp d
y}\overset{\text{CO}}=&\frac{\alpha_s (\pi \bar{R}_A^2)}{(2\pi)^{7}
(N_c^2-1)} \underset{\vka,\vk}{\int}
\frac{\varphi_{p,y_p}(\vka)}{k_{1\perp}^2}\\
&\times \mathcal{N}_Y(\vk)\mathcal{N}_Y(\vp-\vka-\vk)
\,\Gamma^\kappa_8,
\end{split}
\end{align}
for the color-octet channels. Here $\varphi_{p,y_p}$ is the unintegrated gluon distribution inside the proton, which can be expressed as
\begin{align}\label{eq:unintegrated}
\varphi_{p,y_p}(\vka)=\pi \bar{R}_p^2 \frac{N_c k_{1\perp}^2}{4\alpha_s} \widetilde{\mathcal{N}}^A_{y_p}(\vka)\,.
\end{align}
The functions ${\cal G}^\kappa_1$ ($\Gamma^\kappa_8$) are calculated perturbatively--the expressions are available in \cite{Ma:2014mri} (\cite{Kang:2013hta}).
${\cal N}$ ($\widetilde{\mathcal{N}}^A$) are the  momentum-space dipole forward scattering amplitudes with Wilson lines in the fundamental (adjoint) representation, and $\pi \bar{R}_p^2$ ($\pi \bar{R}_A^2$) is the effective transverse area of the dilute proton (dense nucleus). These formulas can be used to compute quarkonium production in p+A collisions. By replacing ``$A$'s by $p$'s", they can also be used to compute quarkonium production in  p+p collisions~\cite{Ma:2014mri}.
For deuteron-gold (d+Au) collisions at RHIC, since gluon shadowing effects are weak for deuteron side, we assume $\varphi_{d,y_d}(\vka)= 2\, \varphi_{p,y_p}(\vka)$.

Before we confront our framework to data on p+A collisions, there are a number of parameters that have to be fixed. Nearly all the parameters are identical to those previously determined in our study~ \cite{Ma:2014mri} of p+p collisions. {The  charm quark mass is set to be $m=1.5\gev$, approximately one half the $J/\psi$ mass.
The CO LDMEs were extracted in the NLO collinear factorized NRQCD formalism~\cite{Chao:2012iv} by fitting Tevatron high $p_\perp$ prompt $\jpsi$ production data; one obtains  $\mops=1.16/(2N_c) \gev^3$, $\mopa=0.089\pm0.0098 \gev^3$, $\mopb=0.0030\pm0.0012 \gev^3$ and $\mopc=0.0056\pm0.0021\gev^3$.
We emphasize, as previously, that the high sensitivity of short distance cross-sections to quark mass
can be mitigated by the mass dependence of the LDMEs.  Note that the uncertainties of these CO LDMEs include only uncorrelated statistic errors, but not correlated errors \cite{Chao:2012iv}.}
Further, as in \cite{Ma:2014mri}, ${\cal N}$ and $\widetilde{\mathcal{N}}^A$ are obtained by solving the running coupling Balitsky-Kovchegov
(rcBK) equation~\cite{Balitsky:1995ub,Kovchegov:1999yj} in momentum space with McLerran-Venugopalan (MV) initial conditions~\cite{McLerran:1993ni,McLerran:1993ka} for the dipole amplitude at the initial rapidity scale $Y_0\equiv\ln(1/x_0)$ (with $x_0=0.01$) for small $x$ evolution. In the case of p+p collisions, all the parameters in the rcBK evolution are fixed from fits to the HERA DIS data~\cite{Albacete:2012xq}. In \cite{Ma:2014mri}, we devised a matching scheme that allowed us to interpolate between the proton's collinearly factorized gluon distribution at large $x$ with the unintegrated distribution in Eq.~(\ref{eq:unintegrated}). This allowed us to fix the remaining free parameter, the effective gluon  radius of the proton $\bar{R}_p=0.48$ fm.

Turning to p+A collisions, there are two additional parameters in our framework, the initial saturation scale $Q_{s0,A}$ in the nucleus and the effective transverse radius $\bar{R}_A$. The latter is not the charge radius of the nucleus, but parametrizes the overall non-perturbative cross-section of relevance to quarkonium production. A more detailed treatment would take into account the impact parameter dependence of the unintegrated distributions, and model the inelastic proton-nucleus cross-section as in \cite{Miller:2007ri,d'Enterria:2003qs}. We will return to this point shortly.
In general, we can express the initial saturation scale in the nucleus as $Q_{s0,A}^2=N\times Q_{s0, p}^2$, where $N$ is a number to be determined and $Q_{s0, p}^2$ is the initial saturation scale in proton, fixed by the fit to HERA DIS data~\cite{Albacete:2012xq}. Good fits to extant electron-nucleus (e+A) DIS data were obtained in \cite{Dusling:2009ni} for  rcBK evolution with the following initial conditions:
i) MV model with anomalous dimension $\gamma=1.13$,  ii) MV model with anomalous dimension $\gamma=1$. For the initial conditions i), one obtains a good fit to e+A data for $N\approx 3$,
while for initial conditions ii), $N\approx 1.5$. In this paper, rcBK evolution for nuclei was performed for initial conditions ii). To avoid fine tuning, we will choose $N=2$ for the results presented in this paper\footnote{The quality of fit for $N=2$ is marginally better than that for $N=3$ but significantly better than those for $N=1$ or higher values of integer $N$. For the IP-sat model, for median impact parameters in e+A DIS, $N=6$ in contrast to $N\approx 2$ for $b=0$~\cite{Kowalski:2007rw}.}.

Similar to $\bar{R}_p$ for the proton, the effective radius $\bar{R}_A$ providing the non-perturbative normalization of the cross-section here can be different from the  transverse charge radius of the nucleus  because we have a specific heavy particle produced in the final state. Fortunately, there is a physical condition which we can use to constrain it. When $p_\perp$ is much larger than the saturation scale involved, the gluon distribution becomes dilute and the nuclear suppression effect should be negligible. Thus $R_{pA}$ must approach 1 for high $p_\perp>> Q_{s0,A}$. Using Eqs.~(\ref{eq:RpA})-(\ref{eq:unintegrated}) one can derive (the argument is presented in Appendix), the expression,
\begin{align}\label{eq:RA}
\frac{\bar{R}_A^2}{A \bar{R}_p^2}\frac{ Q_{s0,A}^{2\gamma}}{Q_{s0,p}^{2\gamma}}\approx1.
\end{align}
We will see later that Eq.~\eqref{eq:RA} indeed guarantees $R_{pA}\to 1$ at high $p_\perp$ limit, within a few percent. By choosing $\gamma=1$ and $N=2$, we obtain $\bar{R}_A=\sqrt{A/2} \bar{R}_p$, which equals to $4.9\mathrm{~fm}$ for Pb and $4.8\mathrm{~fm}$ for Au\footnote{Interestingly, the ratio $\bar{R}_A/\bar{R}_p\sim 10$ here is close to the ratio of radii  extracted from estimates of the inelastic p+A and p+p cross-sections at both LHC and RHIC.}.

Because Eqs.~\eqref{eq:dsktCS}-\eqref{eq:unintegrated} are computed only at LO in the CGC power counting, the CGC+NRQCD framework cannot be extended to describe high $p_\perp$ p+p and p+A data, one might challenge that using Eq.~(\ref{eq:RA}) to determine $\bar{R}_A$ is not especially meaningful. We emphasize however that this condition must be satisfied for the CGC+NRQCD framework to be self-consistent at each order in the perturbative expansion. The $p_\perp$ at which Eq.~(\ref{eq:RA}) is saturated may differ. At NLO, the above procedure should be redone to determine a new self-consistent condition.


\begin{figure}[!htbp]
\center{
\includegraphics*[scale=0.4]{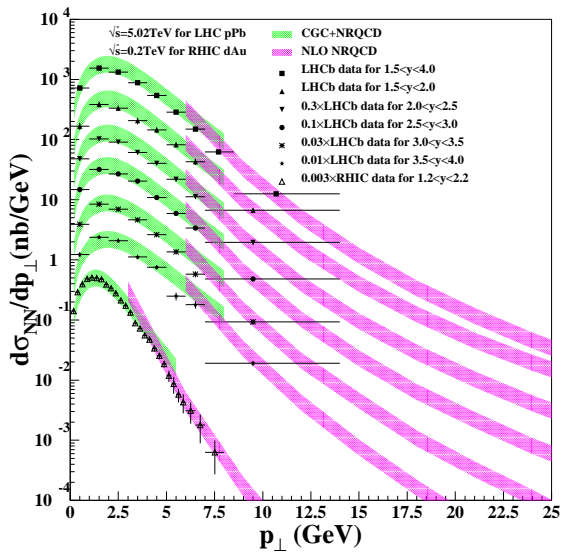}
\caption{\label{fig:pt-pa} $p_\perp$ spectrum of $J/\psi$ production in p+Pb collisions at 5.02 TeV and d+Au collisions at 0.2 TeV. NLO NRQCD results are taken from Ref. \cite{hfzhang2015}. The experimental data are taken from Refs.~\cite{Aaij:2013zxa,Adare:2012qf}.}
}
\end{figure}

To better present the p+A results, we define a cross section per nucleon-nucleon collision, $d\sigma_{NN}=\frac{d\sigma_{AB}}{AB}$. Fig.~\ref{fig:pt-pa} displays the $p_\perp$ spectrum of $J/\psi$ production in p+Pb collisions at 5.02 TeV and d+Au collisions at 0.2 TeV. The bands of our CGC+NRQCD results estimate uncorrelated errors of LDMEs and an additional global $30\%$ uncertainty to account for correlated errors of LDMEs, errors from treatment of feed down, velocity corrections and radiative corrections. We find that the contribution of the CS channel is about $15-20\%$ at small $p_\perp$ and decreases as $p_\perp$ becomes larger.  The NLO NRQCD predictions are taken from \cite{hfzhang2015}, where the PDF shadowing model EPS09~\cite{Eskola:2009uj} was employed to estimate the (small) nuclear shadowing effects at large $p_\perp$. For all rapidity bins available,
the CGC+NRQCD curves match on to the NLO NRQCD ones smoothly, providing a good description of all experimental data. Interestingly, one finds that the CGC+NRQCD curves overshoot the data at smaller values of $p_\perp$ at RHIC relative to the LHC data.  This may be anticipated because, for a given $p_\perp$, small $x$ logs are less important at lower energies. However, a full NLO computation in this framework is needed to understand better the matching in $p_\perp$ of the two formalisms.

\begin{figure}
\center{
\includegraphics*[scale=0.4]{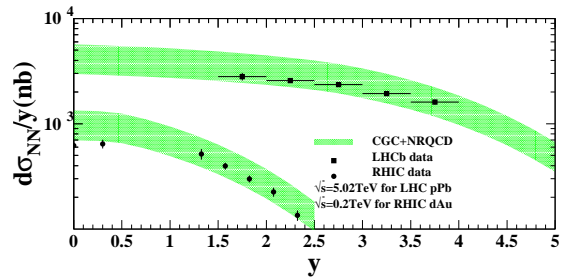}
\caption{\label{fig:y-pa} Rapidity distribution of $J/\psi$ production in p+Pb collisions at 5.02 TeV and d+Au collisions at 0.2 TeV.
The experimental data are taken from Refs.~\cite{Aaij:2013zxa,Adare:2010fn}.}
}
\end{figure}

The rapidity distribution of $J/\psi$ production in p+Pb collisions at 5.02 TeV and d+Au collisions at 0.2 TeV is shown in Fig.~\ref{fig:y-pa},
where the the bands are generated similarly to those in Fig.~\ref{fig:pt-pa}. Since these data are integrated over $p_\perp$, the low $p_\perp$ region dominates and the CGC+NRQCD formalism at LO should apply.  Both LHC data and forward RHIC data are well covered by our uncertainty band; the central value for mid-rapidity RHIC data however is slightly below the band. For this data point, our theory curves should have a larger systematic uncertainty because our framework is most reliable for dilute-dense collisions corresponding to high energies and forward rapidities. The key observation though is that both the relative shapes  as well as the absolute magnitudes of the curves are well captured in the CGC+NRQCD formalism.
The quality of the fits to the $p_\perp$ and rapidity spectra in Figs.~\ref{fig:pt-pa} and \ref{fig:y-pa} are similar to those in p+p collisions~\cite{Ma:2014mri}. Thus we should be able to describe the  $R_{pA}$ ratio, which we shall now discuss.


A key point is that the the large uncertainties for LDMEs, feed down contributions and velocity corrections, largely cancel in the ratio of each NRQCD channel contributing to $J/\Psi$ production. The band spanned by different channels should be able to bracket the $R_{pA}$ value for $\jpsi$ production. With this method, the bounded value of $R_{pA}$ extracted for $\jpsi$ production is independent of the LDMEs and their statistical uncertainties. This is especially noteworthy since independent extractions of the LDMEs from present data are not feasible; their magnitudes, especially between the various CO channels, can vary significantly. Finally, since the CEM is a special case of NRQCD with the  choice of certain LDMEs \cite{Bodwin:2005hm}, our calculation of $R_{pA}$ will also cover the range of CEM predictions. In this sense, the range of theoretical estimates of $R_{pA}$ for $J/\psi$ production are independent of the $\jpsi$ hadronization model and are directly sensitive to the short distance physics.

We will employ here the principal channels for $\jpsi$ production given by NRQCD power counting--these correspond to the $\CScSa$, $\COaSz$, $\COcSa$ and $\COcPj$ channels. \begin{figure}
\center{
\includegraphics*[scale=0.4]{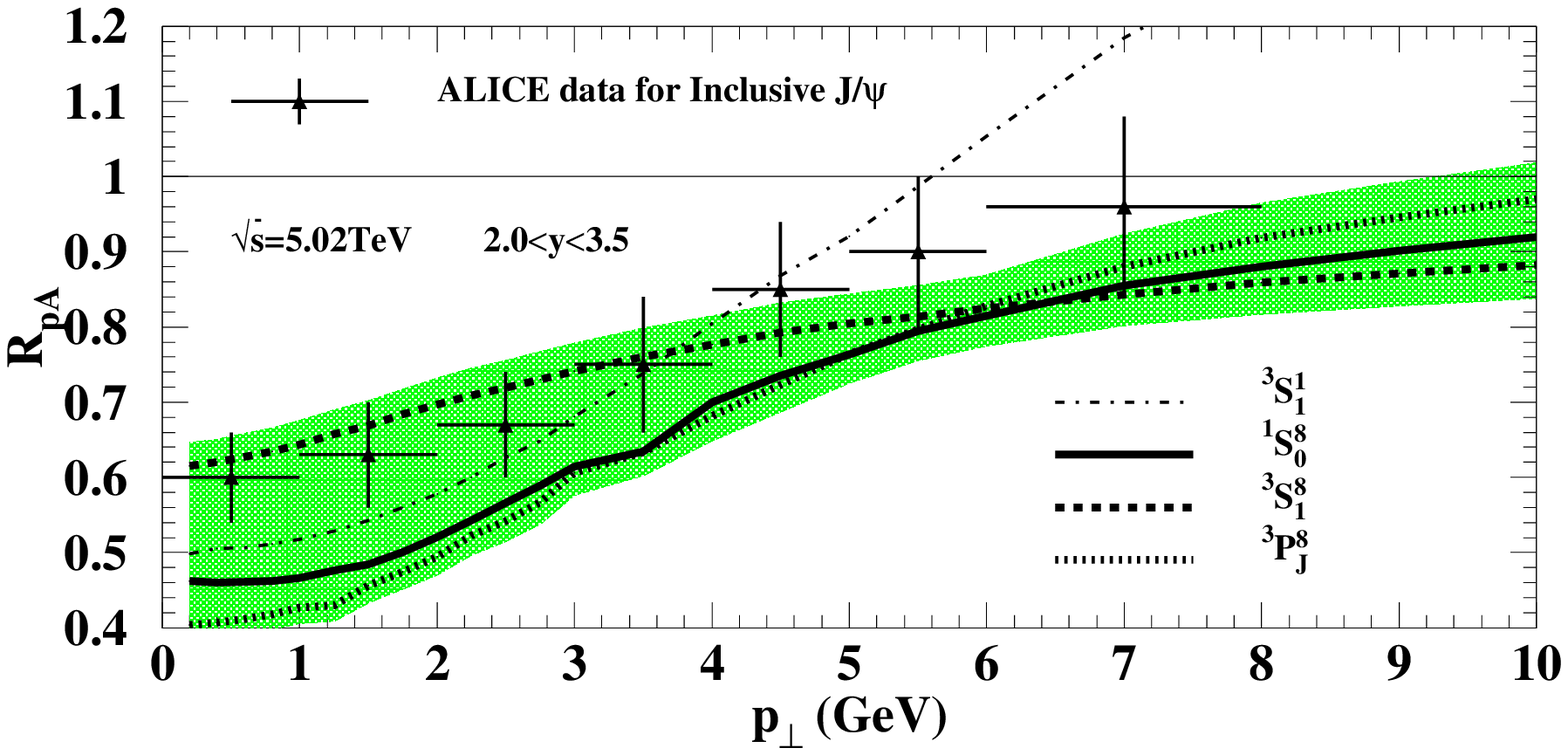}\\
\includegraphics*[scale=0.4]{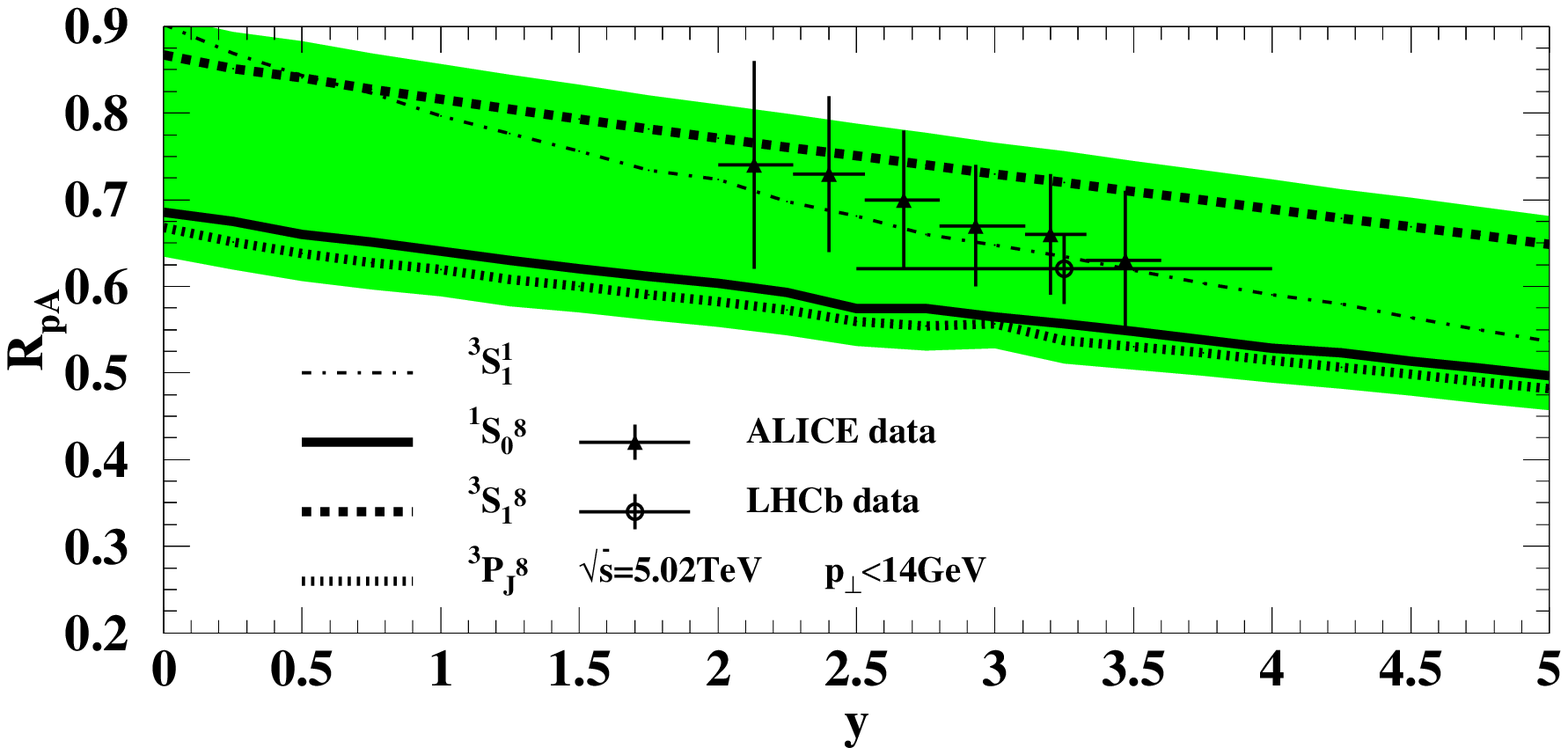}
\caption{\label{fig:rpa-alice} $R_{pA}$ as a function of $p_\perp$ (upper) and rapidity (lower) at LHC.
The experimental data are taken from Refs.~\cite{Adam:2015iga, Abelev:2013yxa,Aaij:2013zxa}.}
}
\end{figure}
\begin{figure}
\center{
\includegraphics*[scale=0.4]{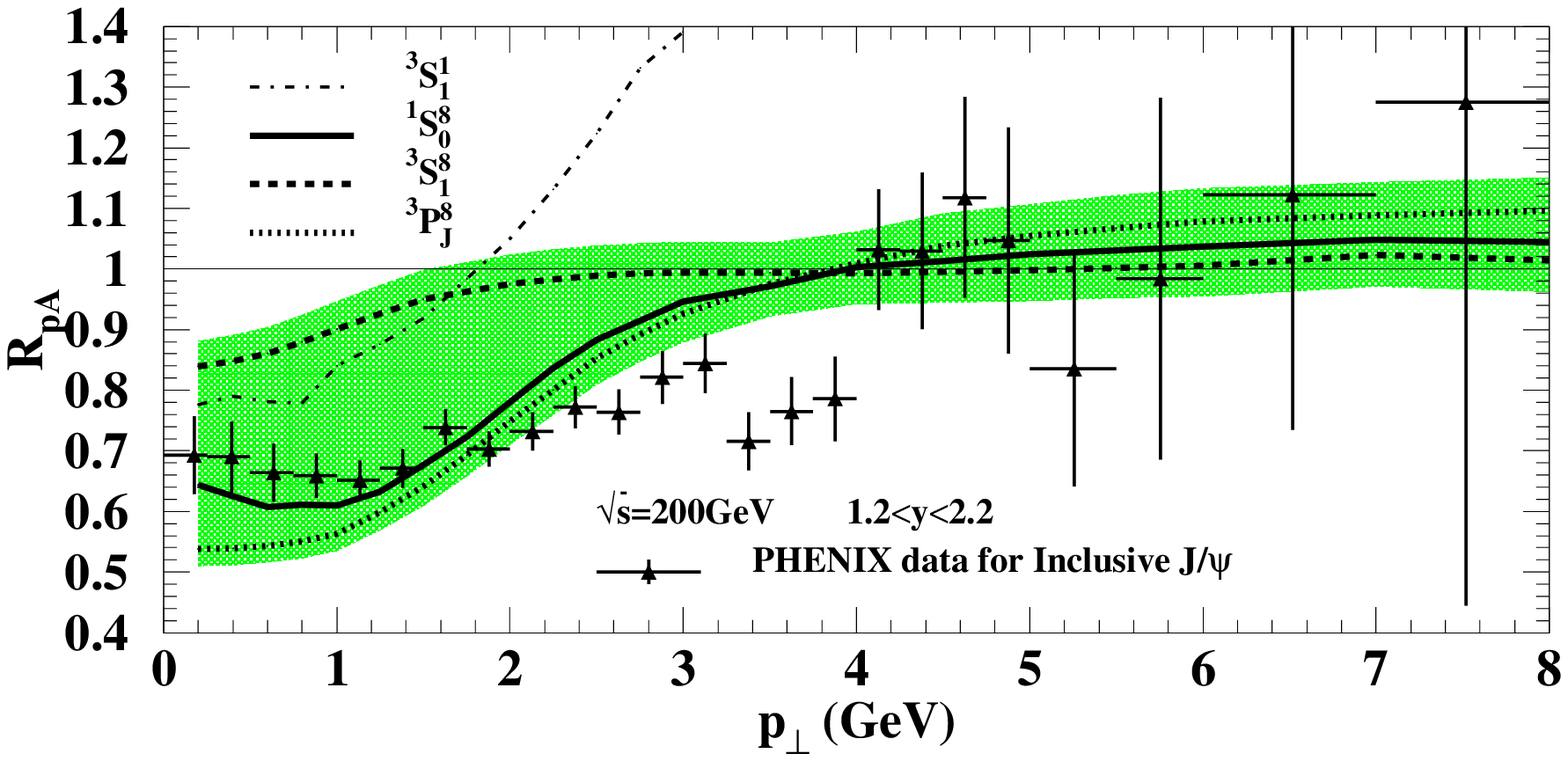}\\
\includegraphics*[scale=0.4]{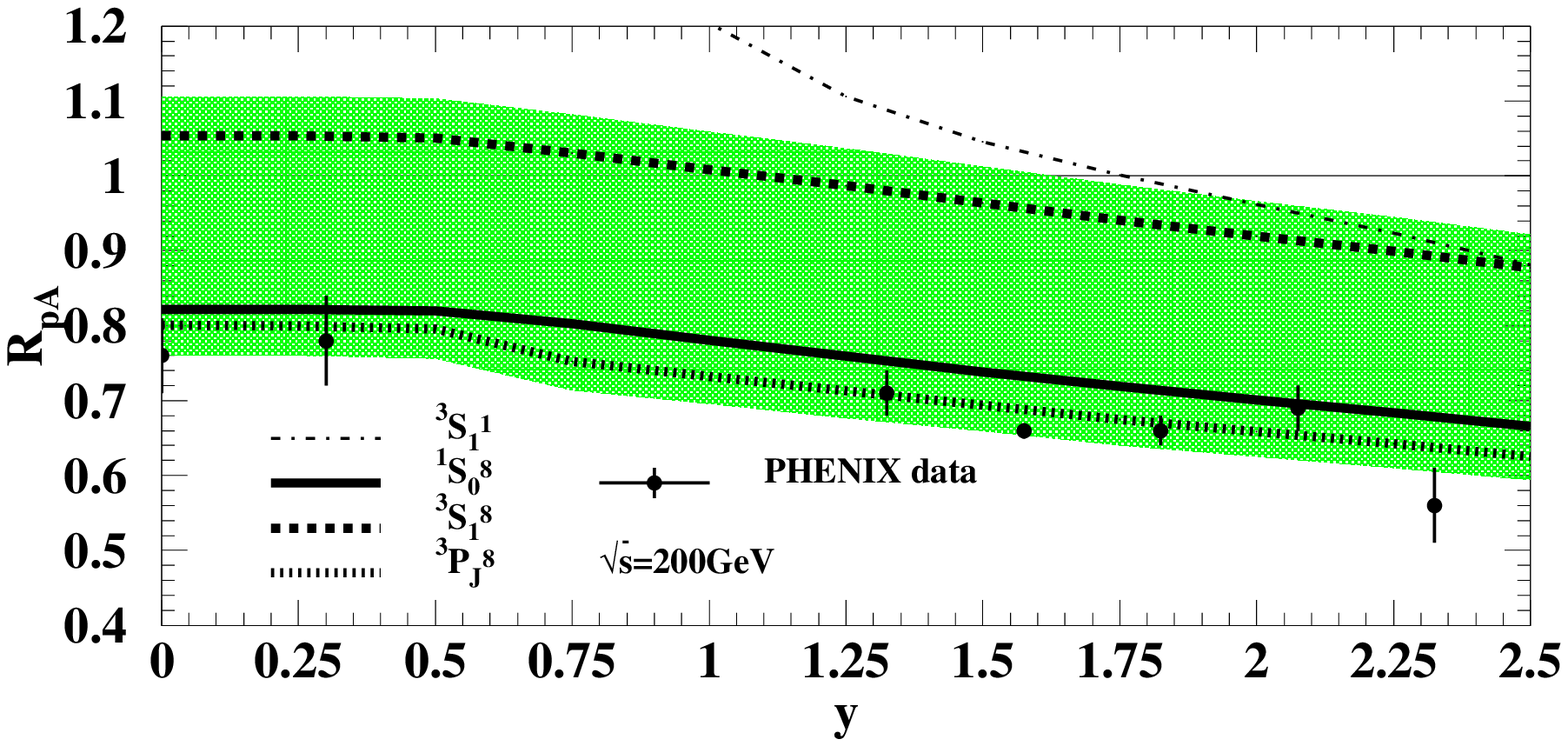}
\caption{\label{fig:rpa-rhic} $R_{pA}$ as a function of $p_\perp$ (upper) and rapidity (lower) at RHIC.
The experimental data are taken from Refs.~\cite{Adare:2012qf, Adare:2010fn}.}
}
\end{figure}
Our results for $R_{pA}$ as a function of $p_\perp$ and rapidity, compared to data from the LHC and  RHIC, respectively,  are presented in Figs.~\ref{fig:rpa-alice} and \ref{fig:rpa-rhic}, where a 5\% systematical error is assumed for each channel to account for the approximation in Eq.~\eqref{eq:RA}. The $R_{pA}$ of all CO channels approaches 1 at high $p_\perp$, confirming that condition Eq.~\eqref{eq:RA} indeed is satisfied by the full theoretical calculation. On the contrary, $R_{pA}$ of the CS channel $\CScSa$ increases to be larger than 1 at high $p_\perp$.  Since forming a color singlet requires two gluons from the target, the additional gluon exchange from the nucleus, at high $p_\perp$, is enhanced relative to that from a proton (by an amount that is proportional asymptotically to the ratio of their saturation scales at the rapidity of interest). Nevertheless, as we find the contribution of the CS channel is small relative to the CO terms in both p+p and p+A collisions, it does not affect our estimate of $R_{pA}$. Thus the band representing the $R_{pA}$ spanned by the CO channels corresponds to our result for $R_{pA}$ of $J/\psi$ production.

The $p_\perp$ and rapidity $R_{pA}$ data from both RHIC and LHC lie within our uncertainty bands.
At the LHC, the $\COcSa$ state lies closest to the central values of the data, while at RHIC, the  $\COaSz$ and $\COcPj$ channels are closest to the data.  Our results suggest that the $R_{pA}$ data, in a future global analysis within the CGC/NLO+NRQCD framework,  can help constrain the LDMEs more stringently, thereby providing a further test of NRQCD.

To summarize, we have shown here that $J/\psi$ spectra in p+A collisions both at RHIC and the LHC are well described by our CGC+NRQCD computations. The two free non-perturbative parameters are related by Eq.~\eqref{eq:RA}; further, the value of the initial nuclear saturation scale $Q_{s0,A}$ is consistent with the values that best describe fixed target e+A DIS data. The fact that the $R_{pA}$ $p_\perp$ data lie within the bands spanned by our computations for the different color octet channels is a strong evidence for the robustness of our framework since these curves are insensitive to details of how heavy quark pairs hadronize to form the $J/\psi$. The results in this paper, when combined with those in \cite{Ma:2014mri}, provide the first comprehensive description of $J/\psi$ production in both p+p and p+A collisions at collider energies.

Several outstanding questions remain. Firstly, the NLO CGC computation needs to be performed to confirm that the framework established is solid. Secondly, other quarkonium states remain to be studied; these come with unique challenges. For instance, for $\Upsilon$ production, Sudakov type double logs in $M/P_\perp$ are important and need to be resummed~\cite{Berger:2004cc,Sun:2012vc,Qiu:2013qka}. A systematic computation of $\psi(2S)$ production in p+A collisions, may require that we include relativistic contributions in the computation of the heavy quark matrix elements. All these questions can be explored in the framework discussed here.

We thank Roberta Arnaldi and Prithwish Tribedy  for helpful communications. This work was supported in part by the U.S. Department of Energy Office of Science under Award Number DE-FG02-93ER-40762, U. S. Department of Energy under Contract No.  de-sc0012704, and the National Natural Science Foundation of China No. 11405268.

\appendix

\section{Derivation of Eq.~(\ref{eq:RA})}

Let us derive the corresponding relation from this condition. As CS contribution is negligible at very high $p_\perp$ regime \cite{Kang:2013hta}, we will only consider the CO contribution in Eq.~\eqref{eq:dsktCO}. When $p_\perp$ is large, at least one of $\vka$, $\vk$ and $\vp-\vka-\vk$ needs to be large. As we are considering a dilute-dense collision, the contribution from the region where $\vka$ is large should be less important, which implies we can take the collinear limit for the proton side and give \cite{Kang:2013hta}
\begin{align}\label{eq:dscollCO}
\frac{d \hat{\sigma}^\kappa}{d^2\vp d
y}\overset{\text{CO}}\approx&\frac{\alpha_s (\pi \bar{R}_A^2)}{4(2\pi)^{3}
(N_c^2-1)} {x_p f_{p/g}(x_p,Q^2)} \nonumber \\
&\times \underset{\vk}{\int}
\mathcal{N}(\vk)\;\mathcal{N}(\vp-\vk) \;\tilde{\Gamma}^\kappa_8\,,
\end{align}
where $f_{p/g}$ is the gluon collinear PDF and $\tilde{\Gamma}^\kappa_8$ are collinear limit of ${\Gamma}^\kappa_8$ which have been calculated in \cite{Kang:2013hta}. Because $\mathcal{N}(\vk)$ decreases as inverse powers of $k_\perp$, the dominant contribution for Eq.~\eqref{eq:dscollCO} comes from two regions, either $\vk$ is small or $\vp-\vk$ is small. It is clear that $\tilde{\Gamma}^\kappa_8$ must be a symmetric function under the transformation $\vk \to \vp-\vk$, thus, in either case, we can set $\vk$ to be zero in $\tilde{\Gamma}^\kappa_8$ and it becomes independent of $\vk$. Then we can perform the $\vk$ integration in Eq.~\eqref{eq:dscollCO}, which gives $\widetilde{\mathcal{N}}^A_{Y_A}(\vp)$. Therefore, the $R_{pA}$ defined in Eq.~\eqref{eq:RpA} behaves as
\begin{align}\label{eq:RpAlimit}
R_{pA}\overset{\text{high $p_\perp$}}\longrightarrow \frac{\bar{R}_A^2}{A \bar{R}_p^2}\frac{ \widetilde{\mathcal{N}}^A_{Y_A}(\vp)}{\widetilde{\mathcal{N}}^A_{Y_p}(\vp)}.
\end{align}
Note that, although we have used the collinear approximation to derive the above relation, the relation holds much better than the collinear approximation itself, which is caused by the cancellation between the contributions to the numerator and denominator of $R_{pA}$ from large $\vka$ region. It is known that MV model with rcBK equation gives $\widetilde{\mathcal{N}}^A_{Y_A}(\vp) \propto Q_{s,A}^{2\gamma}$ at high $\vp$ limit \cite{Iancu:2003xm}, we therefore have $\frac{ \widetilde{\mathcal{N}}^A_{Y_A}(\vp)}{\widetilde{\mathcal{N}}^A_{Y_p}(\vp)}\approx\frac{ Q_{s,A}^{2\gamma}}{Q_{s,p}^{2\gamma}} \approx \frac{ Q_{s0,A}^{2\gamma}}{Q_{s0,p}^{2\gamma}}$, where at the last step we assume that the $Y$ is not significantly larger than $Y_0$ and thus the ratio of saturation scales is not changed too much by evolution. Following these steps, we obtain the condition in Eq.~(\ref{eq:RA}).

\providecommand{\href}[2]{#2}\begingroup\raggedright\endgroup

\end{document}